\documentclass[12pt]{article}
\usepackage{graphicx}

\newsavebox{\sboxpubnumber}
\newsavebox{\sboxpubdate}
\newcommand{\pubdate}[1]{\begin{lrbox}{\sboxpubdate}{#1}\end{lrbox}}

\newcommand{\Title}[1]{\begin{center} {\Large #1 } \end{center}}
\newcommand{\Author}[1]{\begin{center}{ \sc #1} \end{center}}
\newcommand{\Address}[1]{\begin{center}{ \it #1} \end{center}}
\newcommand{\andauth}{\begin{center}{and} \end{center}}

\newenvironment{Abstract}{\begin{quotation}  }{\end{quotation}}
\newenvironment{Presented}{\begin{quotation} \begin{center}
             PRESENTED AT\end{center}\bigskip
      \begin{center}\begin{large}}{\end{large}\end{center}
      \end{quotation}}

\begin{document}

\begin{titlepage}
\pubdate{\today}                    

\vfill
\Title{Baryogenesis in models with large extra dimensions}
\vfill
\Author{Anupam Mazumdar}
\Address{The Abdus Salam International Centre for Theoretical Physics, \\
         I-34100, Italy}
\vfill
\andauth
\vfill
\Author{Rouzbeh Allahverdi~$^{1}$, Kari Enqvist~$^{2}$, and
Abdel P\'erez-Lorenzana~$^{3}$ }
\Address{$^{1}$ Physik Department, TU Muenchen, James Frank Strasse, \\
D-85748, Garching, Germany. \\
$^{2}$ Department of Physics and Helsinki Institute of Physics,
P. O. Box 9, FIN-00014, University of Helsinki, Finland.\\
$^{3}$ The Abdus Salam International Centre for Theoretical Physics,\\
 I-34100, Trieste, Italy.\\
$^{4}$ Departamento de F\'{\i}sica,
Centro de Investigaci\'on y de Estudios Avanzados del I.P.N.\\
Apdo. Post. 14-740, 07000, M\'exico, D.F., M\'exico.}

\vfill
\begin{Abstract}
We describe how difficult it is to realise baryogenesis in models where
the fundamental scale in nature is as low as TeV. The problem becomes 
even more challenging if we assume that there are only {\it two} extra 
compact spatial dimensions, because thermal history of such a Universe 
is strongly constrained by various cosmological and astrophysical bounds 
which translate the maximal temperature of the Universe which must not 
exceed $\sim {\cal O}(10)$MeV. This simply reiterates that the observed 
baryon asymmetry must be synthesised just above the nucleosynthesis 
scale. In this talk we address how to construct a simple model which can 
overcome this challenge. 
\end{Abstract}
\vfill
\begin{Presented}
    COSMO-01 \\
    Rovaniemi, Finland, \\
    August 29 -- September 4, 2001
\end{Presented}
\vfill
\end{titlepage}
\def\thefootnote{\fnsymbol{footnote}}
\setcounter{footnote}{0}

\section{Introduction}

Recently it has been proposed that large extra spatial
dimensions can explain the apparent weakness of the
electroweak scale compared to gravity in $3+1$ dimensions.
In such a scenario four dimensional world is assumed to be a flat
hypersurface, called a `brane', which is embedded in a higher dimensional
space-time usually known as the bulk. The hierarchy
problem is then resolved by assuming that TeV scale can  be
the fundamental scale in higher dimensions~\cite{nima0,early}.
This however requires the compactified radii of the extra dimensions 
are large. Such a large volume can substantiate the hierarchy 
in energy scales. The volume suppression $V_{d}$, the effective 
four dimensional Planck mass $M_{\rm p}$, and, the fundamental 
scale in $4+d$ dimensions $M_{*}$ are all related to each other 
by a simple relation
\begin{equation}
M_{\rm p}^2 =M_{*}^{2+d}V_{d}\,.
\end{equation}
This automatically sets the present {\it common size} of all the extra
dimensions at $b_0$. For two extra dimensions, and, $M_{*}=1$ TeV, the required
size is of order $0.2$ mm right on the current experimental limit for the
search of deviation in Newtons gravity~\cite{exp2}.  Recent astrophysical
and cosmological bounds suggest $M_\ast$ should be larger than $500$ TeV
for two extra dimensions, and around $30$ TeV for three extra dimensions
\cite{han}. Naturally, such a model has an important impact on collider 
experiments \cite{exp1}, and on cosmology (see for instance Ref.~\cite{kari}
and references therein). In this talk we 
address one of the most important question of cosmology and particle 
physics; how to realise the observed baryon asymmetry $\sim 10^{-10}$ in 
nature ? 

\vskip7pt

The generation of baryon asymmetry requires three well-known conditions;  
$C$ and $CP$ violation, $B$ and/or $L$ violation, and out of equilibrium 
decay or scattering processes \cite{sakh}. It is quite likely that the 
departure from thermal equilibrium is possible at early times when 
the expansion rate of the Universe is large. Certainly acquiring such a 
condition becomes more difficult at late times, especially if thermal 
equilibrium is reached at a temperature below the electroweak scale.
Note that the size of the compact extra dimensions must be stabilised
and the corresponding mass scale for two extra dimensions is of order 
$\sim 10^{-2}$ eV. Above this scale the Kaluza-Klein (KK) modes of gaviton 
can be excited very easily, and as we reach higher in temperature we
excite them more and more. We must note that the bulk is also very large 
which can dilute their number density, and you might suspect that we 
would never be able to feel their presence. However, this is not correct
because these KK modes can decay into light fermions via Planck suppressed
couplings. The readers must note that the KK modes once produced can
never reach thermal equilibrium and their number density is frozen.
Therefore, the KK modes when decay their decay products can have an 
energy scale proportional to their mass, which could be related to the 
largest temperature of the thermal bath produced in the early Universe.

\vskip7pt

One could envisage that the largest cross section of a heavy KK mode
is possible at a centre of mass frame; $\gamma+\gamma\rightarrow G$,
where $\gamma$ denotes the relativistic species and $G$ signifies the 
KK mode. The cross section goes as 
$\sigma_{\gamma + \gamma \rightarrow G} \sim (TR)^{d}/M_{\rm p}^2$
\cite{nima0}, where $R$ is the effective size of the extra dimensions.
Once the KK modes are produced their evolution can be traced and a simple
overclosure of their density; $n_{\rm G}/n_{\gamma} \leq 1$ limits the 
initial thermal bath of the Universe which we denote here by $T_{\rm r}$,
usually known as the largest temperature during the radiation dominated era.
\begin{equation} 
\label{final1}
T_{\rm r} \leq T_{\rm c} 
\sim \left(\frac{M_{*}^{d+2}}{M_{\rm p}}\right)^{1/1+d}\,.
\end{equation}
where $T_{\rm c}$ is the normalcy temperature which states that the 
Universe better thermalises below this bound. For some preferred values  
$M_{*} \sim {\cal O}(10)$TeV for $d=2$, we obtain
$T_{\rm c} \leq {\cal O}(10-100)$ MeV. This is an extremely strong 
constraint on thermal history of the Universe. It reiterates very 
strongly that the radiation dominated  Universe simply can not 
prevail beyond this temperature. Above the normalcy temperature our
Universe is inevitably dominated by the KK degrees of freedom.
Therefore, any physical phenomena such as first order phase transition,
out-of-equilibrium decay of heavy particles, if at all taking place beyond
$T_{\rm c}$ must take the KK degrees of freedom into account. 
Note that for larger number of compact extra spatial dimensions the 
bound on normalcy temperature is relaxed quite a lot. Therefore, it is the
two extra dimensions  scenario which poses the most challenging problems
for theorists and also to the experimentalists who are trying their 
luck in seeking any departure from the Newtons gravity below mm. scale
\cite{exp2}.
 
\vskip7pt

Apart from the above mentioned problems there exists another 
major problem with a fast proton decay. The low fundamental scale 
induces proton decay via dimension $6$ baryon number violating 
operator in the theories beyond the Standard Model (SM). With such 
a low fundamental scale the usual coupling suppression is not sufficient. 
We will come back to this issue when we discuss the baryon number violating
lepto-quark interactions which is also required for baryon asymmetry
\cite{kari}.

\vskip7pt

In nut-shell we require an inflationary model which automatically 
provides very low reheat temperature. Such a possibility is not remote,
especially if we assume that the inflaton sector is a SM gauge singlet
which resides in the bulk. After compactification the effective four 
dimensional theory provides a Planck suppressed couplings of the 
inflaton sector to the SM fields very naturally. Keeping inflaton field
in the bulk also aids stabilising the extra dimensions, because at initial
times the natural size of all the dimensions is inevitably the string
scale; $(\rm TeV)^{-1}$. Therefore, not only our three dimensions must 
expand, but also the extra dimensions until they are trapped when their
radii reach of order mm. for instance in the case of two extra dimensions.
The stabilisation process is a dynamical one because the radion field which
determines the size of the extra dimensions obtains a dynamical mass through
its couplings to the trace of the energy momentum tensor of the bulk fields, 
see Ref.~\cite{anu}. Therefore, the radion mass during inflation gets a
correction during inflation of order $\sim H$; the Hubble parameter, 
which makes the radion field heavy and thus its dynamics is frozen and the
radion rolls down to its true minimum very fast in effectively one Hubble 
time \cite{anu}.   

\vskip7pt

The inflaton sector living in the bulk eventually decays after the end 
of inflation, and since the couplings are Planck mass suppressed, the
decay rate of the inflaton into the Higgses, for instance, given 
as~\cite{abdel0,anu,kari}
 \begin{equation}
 \label{decay} 
 \Gamma_{\phi\rightarrow {\rm HH}} \sim {g^2 M_{\ast}^3\over 32 \pi M_P^2}\,,
 \end{equation}
where $g$ is the coupling constant, and $M_{\ast}$ is the fundamental scale.
While deriving the decay rate in Eq.~(\ref{decay}), we have implicitly 
assumed that the mass of the inflaton is roughly the same order as the 
fundamental scale $\sim M$, otherwise the decay rate of the inflaton would be
even more suppressed for smaller mass scale of the inflaton. The estimated 
reheat temperature of the Universe is then given by 
\begin{equation}
T_{\rm r} \sim 0.1\sqrt{\Gamma M_{\rm p}} \sim 1(10) {\rm MeV}\,,
\end{equation}
just right above the temperature required for a successful Big Bang 
nucleosynthesis. Note that this result is independent of the number of 
extra dimensions. This is an another important lesson; once we assume that
the inflaton resides in the bulk a low reheat temperature is a prediction
of the model and then it becomes even more interesting to address the issue
of baryogenesis.

\vskip7pt

At such a low temperature leptogenesis is certainly not possible
because the SM sphaleron transitions which preserves $B-L$ are not 
in equilibrium below $100$ GeV. Therefore, any prior lepton asymmetry 
can not be processed into a net baryon asymmetry. 
Besides this there is a catch in the leptogenesis scenario.
A singlet right handed Majorana neutrino can naturally couple to
the SM lepton doublet, and, the Higgs field in a following way: 
$h \bar{L} H N$. This leads to a potentially large Dirac mass term 
unless the Yukawa coupling $h\sim 10^{-12}$, or, so. Moreover,  
now the see-saw mechanism fails to work, since, the largest  
Majorana mass we may expect can never be larger than  the fundamental 
scale. Therefore, given a neutrino mass  
$\sim h^2 \langle H\rangle^2/M_{\ast} \sim h^2\cdot {\cal O}(1)$ GeV, we  
still have to fine tune $h^2 \leq 10^{-10}$, in order to obtain the 
right  order of magnitude for the neutrino mass. Therefore, the right 
handed neutrinos if they at all exist are more likely to be  residing in
the bulk rather than on the brane. Due to the volume suppression, 
the bulk-brane coupling naturally provides a small coupling~\cite{abdel00}. 
In any case  the  decay rate of the right handed neutrino to the SM fields 
is suppressed by the smallness of $h$ that  gives rise to a decay rate 
which is  similar to Eq.~(\ref{decay}). This makes extremely difficult 
to realise baryogenesis, because eventually when the right handed 
neutrino decays into the SM fields, the background temperature is of 
order of the reheat temperature $\sim {\cal O}(1-10)$~MeV, and, at 
this temperature the sphaleron transition is not at all in equilibrium. 
The sphaleron transition  rate is exponentially suppressed. So, a 
seemingly suitable lepton number might not even get converted to the 
baryons to produce the desired  baryon asymmetry in the Universe. Indeed, 
a larger reheating temperature, at least ${\cal O}(1-100)$ GeV is required 
for making this scenario viable, which is certainly not bad if the number
of extra dimensions are more, because this would relax the normalcy 
temperature. However, the assumption of inflaton being a bulk field might 
have to be judged carefully, because being a bulk field the inflaton 
interaction to the SM fields is always volume suppressed which will 
automatically ensure a low reheat temperature as we have stressed earlier. 

\vskip7pt
 
On the other hand in some cases it is possible that the maximum 
temperature of the Universe is larger than the
final reheat temperature of the Universe. This may happen if the
decay products of the oscillating inflaton are mainly relativistic
species. If the decay products thermalise when the inflaton is still
oscillating and dominating the energy density of the Universe, then
it changes the usual scaling relationship $T\propto a^{-1}$ between 
the temperature and the scale factor. The temperature reaches its maximum
when $a/a_{\rm I} \sim 1.48$, where $a$ denotes the scale factor 
of the Universe and the subscript ${\rm I}$ denotes the era when 
inflation comes to an end. In our case the inflationary scale is 
determined by $H_{\rm I} \sim M_{\ast}$. After reaching the maximum 
temperature, it decreases as
$T \sim 1.3(g_{\ast}(T_{\rm m})/g_{\ast}(T))^{1/4}T_{\rm m}a^{-3/8}$,
where $T_{\rm m}$ denotes the maximum temperature
For $M_{\ast}\sim {\cal O}(10)$ TeV, the maximum temperature could reach 
$T_{\rm m} \sim {\cal O}(10^{5})$ GeV \cite{kolb}. 

\vskip7pt

However, the above mentioned situation might not arise in our case
because the inflaton coupling to the SM particles is extremely weakly coupled
and the comparative decay rates of the inflaton to the Higgses and the SM 
fermions are equally favourable \cite{abdel0,abdel,kari}. Therefore, if 
there were no initial dominance
of relativistic species other than the inflaton, then it is very likely
that the thermalisation is happening very late and the final reheat temperature
is the maximum temperature one could achieve. However, in an extreme case
even if we assume that there exists some initial thermal bath
which could allow sphaleron transitions to occur, then one might imagine
that sphalerons could reprocess a pre-existing charge asymmetry into baryon 
asymmetry~\cite{olive}, which might be reflected in an excess of $e_L$ 
over anti-$e_R$ created during inflaton oscillations. This mechanism 
requires again $(B+L)$-violating processes are out of equilibrium before 
$e_R$ comes into chemical equilibrium, such that the created baryon asymmetry 
could be preserved. Nevertheless, it is important to notice that still 
decaying inflaton field certainly injects more entropy to the thermal bath, 
provided the inflaton dominantly decays into the relativistic degrees of 
freedom. So, an initially large baryon asymmetry has to be created in order
to obtain the right amount of asymmetry just before nucleosynthesis. 
One can easily estimate the amount of dilution that the last stages 
of reheating era will produce. The entropy dilution factor is given by 
\cite{abdel,kari}:
 \begin{equation}
\gamma^{-1}=\left({s(T_{\rm r})\over s(T_{\rm c})}\right)
 =\left({g_{*}(T_{\rm r})\over g_{*}(T_{\rm c})}\right)
 \left({T_{\rm r}\over T_{\rm c}}\right)^3
 \left({a(T_{\rm r})\over a(T_{\rm c})}\right)^3~,
 \end{equation}
where $s$ is the entropy and $T_{\rm c}$ denotes the electroweak temperature
$\sim 100$ GeV. For a low reheat temperature as $T_{\rm r} \sim 1$ MeV, the 
above expression gives rise to $\gamma^{-1} \geq 10^{25}$. While calculating 
the ratio between the scale factors, we have used $T\propto a^{-3/8}$ and 
$g_{*}(T_{\rm c}) \approx g_{*}(T_{\rm r})$. Therefore, including the 
entropy dilution factor we concludes that the initial  $n_{b}/s$ has to 
be extremely large $\geq 10^{15}$ in order to produce the required baryon 
asymmetry at the time of nucleosynthesis, which is $n_{b}/s \sim 10^{-10}$. 
Such a large baryon asymmetry is an extraordinary requirement on any natural
model of baryogenesis, which is almost impossible to achieve in any case. 
 
\vskip7pt

There are couple of important lessons to be learned from the above analysis.
First of all the large production of entropy during the last stages of 
reheating can in principle wash away any baryon asymmetry produced before 
electroweak scale. The second point is that it is extremely unlikely 
that leptogenesis will also work because one needs to inject enough 
lepton asymmetry in the Universe before the sphaleron transitions are 
in equilibrium. The only simple choice  left is to produce directly 
baryon asymmetry, however, just before the end of reheating. The sole 
mechanism which seems to be doing well under these circumstances is 
the Affleck-Dine baryogenesis \cite{AD}, which we shall discuss in the 
following section. 


\section{Affleck-Dine baryogenesis}

A scalar condensate which 
carries non-zero baryonic, or/and leptonic charge survives during inflation
and decays into SM fermions to provide a net baryon asymmetry. 
In our case the AD field; $\chi$, is a singlet carrying some global 
charge which is required to be broken dynamically in order to provide 
a small asymmetry in the current density. This asymmetry can be 
transformed into a baryonic asymmetry by a baryon violating interactions 
which we discuss later on. In order to break this $U(1)_\chi$ charge 
we require a source term which naturally violates $CP$ for a charged
$\chi$ field, and during the non-trivial helical evolution of the 
$\chi$ field generates a net asymmetry in $\chi$ over $\bar\chi$ \cite{abdel}.
This new mechanism has been recently discussed in the context of
supersymmetric inflationary model \cite{zurab}. The $\chi$ field 
obtains a dynamical mass through its coupling to the inflaton sector
which takes place after the end of inflation.

\vskip7pt

We remind the readers that the inflaton energy density must govern the 
evolution of the Universe and the decay products of the inflaton is 
also responsible for reheating the Universe. This happens once
the inflaton decays before $\chi$ decays into SM quarks and leptons.
This decay of $\chi$ via baryon violating interaction generates 
a baryon asymmetry in the Universe which is given by \cite{abdel} 
 \begin{eqnarray}
 \label{ratio}
 \frac{n_{b}}{s} \approx \frac{n_{b}}{n_{\chi}}\frac{T_{\rm r}}
 {m_{\chi}}\frac{\rho_{\chi}}{\rho_{\rm I}}\,.
 \end{eqnarray}
The final entropy released by the inflaton decay is given
by $s \approx \rho_{\rm I}/T_{\rm r}$.  The ratio
$n_{b}/n_{\chi}$ depends on the total phase accumulated
by the AD field during its helical motion in the background of an oscillating 
inflaton field, which can at most be $\approx {\cal O}(1)$.
If we assume that the AD field is a brane-field, then the energy density 
stored in it can at most be: $\rho_{\chi} \approx m^2_{\chi}M_{\ast}^2$. 
On the other hand the energy density stored in the (bulk) inflaton field 
is quite large $\rho_{\rm I}\approx M_{\ast}^2M_{\rm p}^2$, because the 
projected inflaton energy density on the brane has a Planck enhancement
while its couplings to the SM particles has a Planck suppression.
This simply means that the amplitude of the inflaton field is large initially,
which otherwise would not have been possible if the inflaton were a brane 
field. Therefore, the final ratio \cite{abdel}
\begin{equation} 
\frac{n_{b}}{s} \sim \left(\frac{T_{\rm r}}{M_{\rm p}}\right)
\left(\frac{m_{\chi}}{M_{\rm p}}\right)\approx
10^{-34}\left(\frac{m_{\chi}}{M_{\ast}}\right)\ll 10^{-10}\,,
\end{equation} 
for $T_{\rm r} \sim {\cal O}(1-10)$ MeV. 
The conclusion of the above analysis is again disappointing, as it suggests
that the AD baryogenesis also leads to a small $n_b/s$. One way to boost this 
ratio is to assume that the AD field resides in the bulk. In that
case one naturally enhances the ratio $\rho_{\chi}/\rho_{\rm I}$,
however, keeping in mind that it is still less than one in order not to 
spoil the successes of inflation. The projected AD energy density on the 
brane has now a Planck enhancement; $\rho_{\chi} \sim m^2_{\chi}M_{\rm p}^2$.
This reiterates that the initial vacuum expectation value (vev) of the
AD field is quite large, which were not possible in the earlier discussion
where  the AD field was a brane field.  Therefore, the {\it maximum}  
baryon to entropy ratio \cite{abdel}
\begin{eqnarray}
\label{final}
\frac{n_{b}}{s} \approx \left(\frac{T_{\rm r}}{M_{\ast}}\right)
\left(\frac{m_{\chi}}{M_{\ast}}\right)\, \sim 
10^{-10}\left(\frac{m_{\chi}}{{\rm 1 GeV}}\right) ,
\end{eqnarray}
where we have evaluated the right hand side for $T_{\rm r} \sim 10$ MeV and 
$M \sim 10$ TeV. Although, the mass of the AD field requires some 
fine tuning, up to the $CP$ phase, the above ratio can reach the 
observed baryon to entropy ratio quite  comfortably.  Notice, 
that the actual predicted value  also depends on the initial conditions 
on $\chi$ that may render $m_{\chi}$ more freedom.  
Say for instance, if the initial vev of $\chi_0\sim M_{GUT}$, we get the right 
$n_b/s$ provided $m_\chi\sim M$. 

\vskip7pt

We have noticed earlier that due to the violation of $U(1)_\chi$ charge, 
the dynamics of the  AD field generates an excess of $\chi$ over 
$\bar \chi$ fields. This asymmetry is transfered into baryon asymmetry 
by a baryon violating interaction, such as $\kappa\chi Q Q Q L/M^2M_{\rm P}$, 
however, keeping $B-L$ conserved. We also assume that $\chi$ interactions to
SM fields conserve $U(1)_\chi$ symmetry, thus, the 
quarks and leptons must carry a non zero global $\chi$ charge 
while the Higgs field does not. This avoids $\chi$
decaying into Higgses, which otherwise will reduce the baryonic abundance and
make the above interaction the main channel for its decay. 
While discussing the decay rate of $\chi$ field one has to  
take into account all 
possible decay channels which can be of the order of thousands  due to 
family and color freedom. On the other hand, 
we assume that the inflaton is decaying mainly 
into Higgses. Final result is then given by \cite{abdel}
 \begin{equation}
 \Gamma_\chi \approx  \left({\kappa\over g}\right)^2 
 \left({m_\chi\over M_{\ast}}\right)^7~\Gamma_{\phi} ~.
 \end{equation}
By taking $\kappa/g\sim {\cal O}(1)$  we can insure
that $\chi$ will decay along with the inflaton, provided that its mass is 
very close to the fundamental scale. This will certainly demand some level of
fine tuning in the parameters.
We would like to mention that 
this is perhaps the simplest scenario one can think of for generating 
baryon asymmetry right before nucleosynthesis takes place.
It is worth mentioning that in our model 
the AD field will not mediate proton decay by dimension six operators as
$QQQL$, as long as $\chi$ does not develop any vacuum expectation value. 
Notice, other processes mediating proton decay, such  
as instanton effects might still occur.  While there is no known
solution for such a potential problem yet, our mechanism is at least not 
adding any new source to proton decay.
In the same spirit one may check those operators which induce $n-\bar n$
oscillations. Again, effective $\Delta B=2$ 
operators of dimensions 9; $UDDUDD$, and 11; $(QQQH)^2$,
can not be induced by integrating out $\chi$.


\subsection{The model}

In this section we briefly discuss whether we can provide flesh to
the above discussed scenario. Without going into much details we
describe the AD potential which must have to come in conjunction
with the inflaton sector \footnote{In this paper we do not discuss
the details of the inflationary model which can be found in Ref.~\cite{kari}.}.
The AD potential can be given by \cite{kari}
\begin{eqnarray}
\label{adpot}
V_{\rm AD}(\phi,N,\chi_1,\chi_2) =
\kappa_1^2 \left(\frac{M_{\ast}}{M_{\rm p}}\right)^2 N^2(\chi_1^2+\chi_2^2)
 +\frac{\kappa_2^2}{4}\left(\frac{M_{\ast}}{M_{\rm p}}\right)^2
 (\chi_1^2+\chi_2^2)^2\, \nonumber \\
+\kappa_3 ^2\left(\frac{M_{\ast}}{M_{\rm p}}\right)^2\phi N 
(\chi_1^2-\chi_2^2)\,,
\end{eqnarray}
where $\kappa_1,\kappa_2,\kappa_3$ are constants of order one, and 
$\chi_1$ and $\chi_2$ are
the real and imaginary components of the complex field AD field $\chi$. 
For the stability of the potential $\kappa_{1}\sim \kappa_{2}\geq \kappa_{3}$.
Here 
we have two other dynamical fields which we have denoted by $\phi$ and $N$
which are coming from the inflaton sector. Note that the last term in the
above equation which is responsible for breaking global $U(1)_{\chi}$. This
could however be possible if both $\phi,N$ have non zero average vev. It
might be possible to provide a scenario where initially the last term is 
vanishing where there is no apparent violation of a global charge. This 
can be realised in a particular inflationary model where the field is 
trapped in its own local minimum $N=0$ during inflation, while responsible
for generating a large vev which could drive inflation. The role of 
other field $\phi$ is to end inflation via instability through its coupling 
to the $N$ field which we have not written down. This is a perfect example
of hybrid inflationary model where after the end of inflation both $\phi,N$
begin oscillations about their true minimum.  Therefore, in our picture
the first and last term is vanishing during inflation while the AD field 
is massless and its evolution is determined by the quartic term in the
above equation. However, after the end of inflation when both $\phi, N$
oscillates all the terms including the last term in Eq.~(\ref{adpot})
turns on. This is an interesting example of breaking $U(1)_{\chi}$ 
symmetry dynamically towards the end of inflation.

\vskip7pt

Note that in Eq.~(\ref{adpot}), the effective couplings are Planck suppressed,
this is due to the fact that the AD field is a bulk field. If we know the 
post-inflationary behaviour of $\phi$ and $N$ fields, we shall be able to 
estimate the total asymmetry in $\chi$ distribution due to the source term
which is responsible for $CP$ violation, given by the last term in the above
equation. For a charged scalar field this is equivalent to $C$
violation also. The $B$ violation arises via the decay of $\chi$, 
because the decay products have $\Delta B \neq 0$ as discussed in the earlier 
section,  and we have a 
non-trivial helical oscillations in $\chi$ which accumulates net 
$CP$ phase which is transformed into asymmetric $\chi$. The net 
$\chi$ asymmetry; $n_{\chi}$ can be calculated by evaluating the 
Boltzmann equation: 
\begin{eqnarray}
n_{\chi}= \frac{i}{2}\left(\dot \chi^{\ast}\chi-
\chi^{\ast}\dot \chi\right)\,.
\end{eqnarray}
With the help of equations of motion for $\chi_1,\chi_2$ for the potential
Eq.~(\ref{adpot}) we can rewrite the above expression as
\begin{eqnarray}
\label{lep}
\dot n_{\chi}+3Hn_{\chi}=4\kappa_3^2 \left(\frac{M_{\ast}}{M_{\rm p}}\right)^2
\langle N(t)\phi(t)\rangle \chi_1(t)\chi_2(t)\,.
\end{eqnarray}
The right hand side of the above equation is a source term which generates a
net $\chi$ asymmetry through a non-trivial motion of $\chi_1$ and
$\chi_2$ fields. We integrate Eq.~(\ref{lep}) from $t_0$ which corresponds
to the end of inflation up to a finite time interval \cite{kari}.
\begin{eqnarray}
\label{integral}
n_{\chi} a^3 &=& 4\kappa_3^2 \left(\frac{M_{\ast}}{M_{\rm p}}\right)^2
\int_{t_0}^{t}\langle N(t^{\prime}) \phi(t^{\prime})\rangle a^3(t^{\prime})
\chi_{1}(t^{\prime})\chi_{2}(t^{\prime})dt^{\prime}\,.
\end{eqnarray}
The upper limit of integration signifies the end of reheating. We assume that
the oscillations continue until the fields decay completely. Before we perform
the integration, we notice that the integrand decreases in time.  This can be
seen as follows ; first of all notice that the approximate number of
oscillations are quite large before the fields decay 
$\sim (M_{\rm p}/M_{\ast})^2$. This allows us to average $\phi$ and $N$
oscillations. The typical time dependence of oscillating fields follow
$\langle \phi N \rangle \sim \Phi^2(t) \sim 1/t^2$ \footnote{ In the hybrid
inflationary model $N(t) \sim A_0(1+\Phi(t)/3 \cos(m_{\phi}t))$ develops a
vev after the end of inflation, however
$\phi(t) \sim A_0\Phi(t)/(3\sqrt{2})\cos(m_{\phi}t)$ does not, $A_0$ is the 
initial amplitude which is of order $\sim M_{\rm p}$. Such a large amplitude 
is precisely due to the fact that the inflaton sector is in the bulk, and upon 
compactification the field can obtain a large vev with energy density 
$\rho_{\rm I}\sim M_{\ast}^2M_{\rm p}^2$.}. Similarly, the $\chi$ oscillations 
provide $\langle \chi_1 \chi_2 \rangle \sim 1/t^2$ \footnote{We must note that
$\chi$ field must not develop any vev in order to avoid proton decay via
lepto-quark interaction. Therefore 
$\chi_1(t)\sim \chi_2(t)\sim (|\chi(0)|/t) \cos(m_{\chi}t)$.} 
another suppression in time. While taking care of the expansion, where 
the scale factor behaves like 
$a(t)\sim t^{2/3}$ during the inflaton oscillations, the overall
behavior of the integrand follows $\sim 1/t^2$. This suggests that the 
maximum contribution to $\chi$ asymmetry comes only at the initial times 
when $t_0 \sim 1/H_0$, where $H_0 \sim M_{\ast}$ in our case.
The right hand side of the above equation turns out to be \cite{kari}
\begin{equation}
\label{integral1}
n_{\chi} \approx \frac{2}{27}\kappa_3^2 \frac{M_{\ast}^2 |\chi(0)|^2}{H_0}\,.
\end{equation}
We have assumed that the total $CP$ phase, which is given by two factors: an
initial phase determined arbitrarily during inflation and the final dynamical
phase which is accumulated during the oscillations, is of order 
$\sim {\cal O}(1)$. We are also assuming that the coupling 
$\kappa_{3} \sim {\cal O}(1)$.

\vskip7pt

The final ratio of $\chi$ number density produced and the entropy is given by
\cite{kari} 
\begin{eqnarray}
\label{final0}
\frac{n_{\chi}}{s} \approx 
{\cal O}(1) \left({|\chi(0)|\over M_{\rm p}}\right)^2
 \left({T_{\rm r}\over M_{\ast}}\right) 
\end{eqnarray}
where we have used the fact that $s\propto a^{3}$ in our case.
This is the final expression for $\chi$ asymmetry produced during the helical 
oscillations of $\chi$ and this has to be compared with the observed baryon 
asymmetry $\sim n_{B}/s \sim 10^{-10}$. Note that the final expression
depends on the amplitude of $\chi_1 \sim \chi_2 \sim \chi(0)$ at the time when
$U(1)_{\chi}$ symmetry is broken, or, in our case when $\phi$ and $N$ start 
oscillating around their minimum. For a particular example; 
$T_{\rm r} \sim 10$~MeV, and $M_{\ast}\sim 100$~TeV, we require
$\chi(0) \sim 10^{16}$ GeV in order to produce a right magnitude of
baryon asymmetry. As we have discussed earlier the baryon 
asymmetry is injected into thermal bath along with the  inflaton decay
products. It is essential that the thermalisation takes
place after AD field has decayed in order not to wash away the total
baryon asymmetry. 


\section{Conclusions}

In summary, we have presented a natural mechanism of baryogenesis in the      
context of low quantum gravity scale with large compact extra dimensions. Our
mechanism is generic and it is independent of the fundamental scale and the
number of compact extra dimensions.  This mechanism does not rely on any extra
assumption other than invoking a fundamental scalar field that lives in the
$4+d$ dimensional space time.  The baryogenesis scheme can work at any
temperature lower than the electro weak scale because the mechanism does 
not depend on sphaleron transition and relies on baryon violating 
interactions, and therefore does not depend on leptogenesis. The presence
of non-renormalizable couplings inevitably reheats the Universe with a
temperature close to nucleosynthesis, which is really independent of the 
number of extra dimensions. While performing our calculation we have 
implicitly assumed that we had a successful model of inflation which could
also provide the right amount of density perturbations. This seems to be 
still alluding. For a recent work we refer the readers Ref.~\cite{anne}.
The main conclusion of Ref.~\cite{anne} is that it is difficult to realise
a hybrid model which could generate an amplitude for the density perturbations
observed by COBE at a scale larger than the size of the horizon with a 
fundamental scale as low as $\sim 100-500$TeV. It is necessary to push the 
fundamental scale up by $2-3$ orders of magnitude. This certainly relaxes
the stringent bound on normalcy temperature for two extra dimensions.
Therefore, opening up the chances of other mechanisms of baryogenesis to
succeed. In any case the scenario which we have provided here is quite generic
and works independent of any preferred fundamental scale and number of 
extra spatial dimensions.


\end{document}